\documentclass[preprint,prd,showpacs]{revtex4}
\usepackage{graphicx}
\usepackage{eucal}
\usepackage{amsxtra}

\begin{document}
\title{Comment on ``Including Systematic Uncertainties in Confidence Interval 
Construction for Poisson Statistics''}

\newcounter{foots}
\newcounter{notes}

\author{
Gary C.~Hill}\email{ghill@senna.physics.wisc.edu}
\affiliation{
 Department of Physics,
 University of Wisconsin, Madison, WI 53706, USA}


\begin{abstract}
The incorporation of systematic uncertainties into confidence interval
 calculations has been addressed recently in a paper by \hbox{Conrad {\em et al.}} (Physical
Review D {\bf 67} (2003) 012002). In their work, systematic uncertainities in detector
efficiencies and background flux predictions were incorporated following the hybrid 
frequentist-Bayesian
prescription of Cousins and Highland, but using the likelihood ratio ordering
of Feldman and Cousins in order to produce ``unified'' confidence intervals. In 
general, the resulting intervals behaved as one would intuitively expect, i.e.
increased with increasing uncertainties. However, it was noted that for numbers
of observed events less than or of order of the expected background, the intervals
could sometimes behave in a completely counter-intuitive fashion -- being seen to 
initially \emph{decrease} in the face of increasing uncertainties, but only
 for the case of increasing \emph{signal} efficiency 
uncertainty. 
 In this comment, we show that the problematic behaviour is due to 
 integration over  the signal efficiency uncertainty while maximising 
the best fit alternative hypothesis likelihood.
 If the alternative hypothesis likelihood is determined
by unconditionally maximising with respect to both  the unknown signal  and 
signal efficiency uncertainty, 
the limits display the correct intuitive behaviour.
\end{abstract}

\maketitle

In their recent paper, \hbox{Conrad {\em et al.}}\cite{bchd} incorporated uncertainties on signal
and background efficiencies into limit calculations by applying the standard classical
likelihood ratio technique\cite{soa}, recently popularised by Feldman and Cousins\cite{fc98},
to the hybrid Bayesian-frequentist method of Cousins and Highland\cite{ch}. In this method, 
the fixed-but-unknown signal strength parameter $\mu_s$ is treated in a frequentist
fashion, but the unknown experimental efficiencies ($\epsilon_s$ and $\epsilon_b$) are
incorporated by integrating over prior probability distributions $P(\epsilon_s)$ and
$P(\epsilon_b)$. This simplifies the confidence interval calculation by reducing the
dimensionality of the problem down to only one unknown variable. The resulting intervals
showed the correct intuitive behaviour in most cases, i.e. the confidence intervals were
seen to increase as the uncertainties in the efficiencies were increased. However, it
was noted that some counter-intuitive behaviour could occur for cases where the observed
number of events $n_0$ was less than or of order of the expected background $\mu_b$. In
these cases, the {\em limits sometimes initially became  more restrictive as the uncertainties
were increased}, a behaviour which anyone would agree was undesirable. This behaviour was
also noted in an earlier paper by Giunti\cite{giunti}. 
In this comment, we show that
the noted undesirable behaviour is due to the choice of likelihood ratio test implemented
by \hbox{Conrad {\em et al.}}, and show how a choice more consistent with the generalised likelihood
ratio test in the presence of nuisance parameters eliminates this behaviour, leading to intervals
with acceptable behaviour as uncertainties are increased. The key to the correction is the
choice of treatment of the uncertainties in the likelihood ratio denominator; \hbox{Conrad {\em et al.}}
chose to integrate over uncertainties in {\em both} the numerator and denominator, which leads
to the counter intuitive behavior. We show here how integrating in the numerator, but
maximising with respect to the uncertainties in the denominator leads to the correct intuitive
behaviour.

We consider specifically the  problematic case from \hbox{Conrad {\em et al.}}, the determination of a 
confidence limit on an unknown Poisson signal strength $\mu_s$ in the presence of a precisely
known background $\mu_b$ but where the signal efficiency $\epsilon_s$ is also unknown. Before 
considering the approach of \hbox{Conrad {\em et al.}}, we  note that a  completely
frequentist treatment would involve the construction of a confidence region in the \mbox{$\mu_s-\epsilon_s$}
plane, where the acceptability of each possible outcome $n$ under a null hypothesis
\mbox{$\{\mu_s,\epsilon_s\}$} is 
determined by a likelihood ratio test\cite{soa,fc98},
where the likelihood ratio $R$ 
\begin{equation}
\label{clrt}
     R = \frac{\mathcal{L}(n\mid \mu_{s},\mu_{b},\epsilon_s)}
                   {\mathcal{L}(n\mid \hat{\mu}_{s},\mu_{b},{\hat{\epsilon}_s})}
\end{equation}
compares the likelihood under the null hypothesis to the likelihood of the alternative
hypothesis $\{\hat{\mu}_s,\hat{\epsilon}_s\}$ that best describes the possible observation
$n$, i.e. $\hat{\mu}_s$ and $\hat{\epsilon}_s$ are the values that unconditionally maximise the
likelihood of observing $n$ events. Values of $n$ where \mbox{$\mathcal{L}(n\mid \mu_{s},\mu_{b},\epsilon_s)$}
is close to the maximum possible value $\mathcal{L}(n\mid \hat{\mu}_{s},\mu_{b},{\hat{\epsilon}_s})$
are most acceptable and are therefore first included into acceptance intervals during the confidence
interval construction. To reduce the dimensionality of the problem, the suggested procedure\cite{soa}
is to conditionally maximise the numerator with respect to the nuisance parameter $\epsilon_s$ yielding
a likelihood ratio
\begin{equation}
\label{clrt2}
     R = \frac{\mathcal{L}(n\mid \mu_{s},\mu_{b},\hat{\hat{\epsilon}}_s)}
                   {\mathcal{L}(n\mid \hat{\mu}_{s},\mu_{b},{\hat{\epsilon}_s})}
\end{equation}

The Cousins-Highland approach is a hybrid frequentist-Bayesian method
where the  incorporation of a systematic uncertainty in the
efficiency $\epsilon_s$ proceeds by
 integrating over
 a prior probability function \mbox{$P(\epsilon_s\mid \hat{\epsilon}_s,\sigma_{\epsilon_s})$},
 describing our 
knowledge of the nuisance parameter $\epsilon_s$, but where the unknown signal 
strength $\mu_s$ is still treated 
  in a classical frequentist fashion.
 This yields a likelihood function
\begin{eqnarray}
 \mathcal{L}(n\mid \mu_{s},\mu_{b},P(\epsilon_s \mid \hat{\epsilon}_s,\sigma_{\epsilon_s})) = 
  \int_{{\epsilon'_s}} \mathcal{L}(n\mid \mu_{s},\mu_{b},{{\epsilon_s}'})
                   P(\epsilon'_s\mid \hat{\epsilon}_s,\sigma_{\epsilon_s}) \,\mathrm{d} {\epsilon'_s}
\end{eqnarray}
which is used to construct classical confidence intervals 
in the fixed but unknown signal strength $\mu_s$. We still need an ordering principle to
decide which values of $n$ are to be included in the acceptance interval for each null
hypothesis $\mu_s$.  \hbox{Conrad {\em et al.}} used
 a Feldman-Cousins
style likelihood ratio ordering
 principle
(equation \ref{clrt} or \ref{clrt2}) 
to choose which values of $n$ fall into the acceptance interval
for a given $\mu_s$.  This unifies the treatment of upper limits and two-sided confidence
regions. 
 However, it is their
treatment of the uncertainties in the determination of the best alternative hypothesis in
the  likelihood ratio test which leads to the 
counter-intuitive behaviour of the confidence intervals. \hbox{Conrad {\em et al.}} used a likelihood ratio of
the form 
\begin{equation}
\label{CBHDratio}
     R = \frac{\int_{\epsilon_s} \mathcal{L}(n\mid \mu_{s},\mu_{b},{{\epsilon'_s}})
                   P(\epsilon_s '\mid \hat{\epsilon}_s,\sigma_{\epsilon_s}) \,\mathrm{d} {\epsilon'_s}}
                   {{\int_{\epsilon_s} \mathcal{L}(n\mid \hat{\mu}_{s},\mu_{b},{{\epsilon'_s}}) 
                       P(\epsilon'_s\mid \hat{\epsilon}_s,\sigma_{\epsilon_s}) \,\mathrm{d} \epsilon'_s}}
\end{equation}
where the uncertainties are integrated over in both the numerator and denominator.
Integration over the uncertainties has the desired effect on the
numerator  likelihood
where values $n$ near  the mean ${\hat\epsilon_s}\mu_{s} + \mu_{b}$ become less probable in favour of
higher and lower values of $n$ as $\sigma_{\epsilon_s}$ increases.
 Integration over uncertainties while finding $\hat{\mu}_{s}$ still results
in $\hat{\mu}_{s}=0.0$ for $n<\mu_b$. However, for $n>\mu_b$,
 the best fit value $\hat{\mu}_{s}$ for a given $n$ 
as well as the probability of this best fit value used in the denominator
decreases as $\sigma_{\epsilon_s}$ increases. 
The net effect is a shift in the peak of the likelihood ratio distribution to higher $n$,
  that then leads to the problematic behaviour in the
confidence intervals by sometimes shifting the  acceptance region 
\mbox{$\{n_{\mathit{lo}}(\mu_s\mid \sigma_{\epsilon_s}=0.0),n_{\mathit{hi}}(\mu_s\mid \sigma_{\epsilon_s}=0.0)\}$} 
for zero uncertainties to higher values of $n$ where
\begin{equation}
n_{\mathit{lo}}(\mu_s\mid \sigma_{\epsilon_s}>0.0) > 
n_{\mathit{lo}}(\mu_s\mid \sigma_{\epsilon_s}=0.0)
\end{equation}
 as the uncertainty is included. Since
the upper limit $\mu_s^{\mathrm{lim}}$
 for a given observation $n_0$ is the highest value of $\mu_s$ for which $n_0$ is
still in the acceptance interval, shifting the acceptance interval for values of $\mu$ at or below the 
limit  
$\mu_s^{\mathrm{lim}}$ will lead to a 
lowering of the limit. 
An example of this effect
 is shown in table \ref{table}, where we examine the specific
 problematic case from \hbox{Conrad {\em et al.}}. The 90\% confidence level upper
 limits for observations of $n_0=2,4$ and 6 events on
a precisely known background of $\mu_b=6.0$ in the presence of various percentage
 Gaussian uncertainties in the signal efficiency are shown, using  the likelihood ratio
of \hbox{Conrad {\em et al.}} (equation \ref{CBHDratio}).
 We note that our
limits for the \hbox{Conrad {\em et al.}} ordering are not exactly as given in their paper as we
here test the null hypothesis in increments of 0.01, rather than 0.05.
 Nonetheless, the limit for the case of
 $n_0=2$ is seen to decrease as the efficiency error increases from 0 to 40\% $(1.56\rightarrow 1.45)$, before
finally increasing. For the $n_0=4$ and 6 cases, the
limits decrease as $\sigma_{\epsilon}$ goes from 0 to 10\%, but then increase.

We can correct this behaviour by constructing
a hybrid frequentist-Bayesian likelihood ratio test, where the numerator retains the integration over the 
uncertainties but where the alternative hypothesis is found by unconditionally maximising the denominator
as in the pure frequentist tests in equations \ref{clrt} and \ref{clrt2}. This  yields
 the following likelihood ratio 
statistic
\begin{equation}
\label{GCHratio}
     R = \frac{\int_{{\epsilon'_s}} \mathcal{L}(n\mid \mu_{s},\mu_{b},{{\epsilon_s}'})
                   P(\epsilon'_s\mid \hat{\epsilon}_s,\sigma_{\epsilon_s}) \,\mathrm{d} \epsilon_s'}
                  {\mathcal{L}(n\mid \hat{\mu}_{s},\mu_{b},{\hat{\epsilon}_s})}
\end{equation}
where the inclusion into the  acceptance interval of given $n$'s  is
 determined by comparison of their likelihood after integration over uncertainties
to the likelihood of  the best fit $\hat{\mu}_{s}$
 given  no uncertainty.
This way, the likelihoods given any value of the uncertainty $\sigma_{\epsilon_s}$
are always compared to the simple alternative hypothesis \mbox{$\{{\hat{\mu}}_s,\hat{\epsilon}_s\}$} which
best describes the observation $n$, rather to one that changes with
$\sigma_{\epsilon_s}$.

\begin{table}[h]
\vspace{1cm}
\begin{center}
\begin{tabular}{ccccc}
\hline\hline
$n_0$ &$\mu_b$ &signal efficiency & \multicolumn{2}{c}{90\% c.l. upper limits}\\
    &    &uncertainty (\%) &\hbox{Conrad {\em et al.}}  &This work\\ \hline
2     &6.0     &0             & 1.56       & 1.56 \\
      &        &10             & 1.55       & 1.56 \\
      &        &20             & 1.53       & 1.57 \\
      &        &30             & 1.50       & 1.59 \\
      &        &40             & 1.45       & 1.61 \\
      &        &50             & 1.72       & 1.96 \\
4     &6.0     &0             & 2.82       & 2.82  \\
      &        &10             & 2.81       & 2.83 \\
      &        &20             & 3.16       & 3.24  \\
      &        &30             & 3.08       & 3.26 \\
      &        &40             & 3.35       & 3.67   \\
      &        &50             & 3.92       & 4.41 \\
6     &6.0     &0             & 5.46       & 5.46  \\
      &        &10             & 5.43       & 5.46 \\
      &        &20             & 5.74       & 5.89  \\
      &        &30             & 5.97       & 6.31 \\
      &        &40             & 6.51       & 7.12   \\
      &        &50             & 7.63       & 9.03 \\
\hline\hline
\end{tabular}

\caption{\label{table} 
Comparison between upper limits in the presence of signal efficiency uncertainties using
the \hbox{Conrad {\em et al.}} likelihood ratio ordering and the ordering in this
present work. The \hbox{Conrad {\em et al.}} limits initially decrease as uncertainties are increased,
whereas the likelihood ratio ordering described here results in limits with the desired behaviour, i.e.
they increase as the uncertainties increase.
}
\end{center}
\end{table}

Table \ref{table} shows that the limits using the ordering from equation \ref{GCHratio} are seen to initially 
 remain the same (due to overcoverage from the
discrete nature of the Poisson distribution) but then increase with increasing uncertainty.
We also note that although the \hbox{Conrad {\em et al.}} limits do finally increase as the uncertainties
grow bigger, they are always smaller than the limits found using the likelihood ratio
in equation \ref{GCHratio}.

\acknowledgements{
 I wish to thank Jan Conrad for
 useful discussions and Katherine Rawlins for helpful 
comments on the manuscript. This work was supported by the National
Science Foundation under constract number OPP-9980474.}

\end{document}